\documentclass[11pt]{article}
\usepackage{epstopdf}
\usepackage[a4paper,hmarginratio=1:1,vmarginratio=2:3,totalwidth=15.4cm,totalheight=22.7cm]{geometry}
\usepackage{bm,epstopdf,epsfig,amsmath,amssymb,amsfonts,colordvi,latexsym,comment,cancel,
verbatim,
slashed}
\usepackage{graphicx,graphics}
\usepackage[font=md,captionskip=8pt]{subfig}
\usepackage[usenames,dvipsnames]{color}

\usepackage{setspace}

\usepackage[utf8]{inputenc}

\newcommand{\cL}{{\cal L}}

\newcommand{\cO}{{\cal O}}

\newcommand{\cV}{{\cal V}}

\newcommand{\Tr}{\mbox{Tr}}

\renewcommand{\Re}{\text{Re}\ }
\renewcommand{\Im}{\text{Im}\ }

\newcommand{\ra}{\rightarrow}
\newcommand{\be}{\begin{equation}}
\newcommand{\ee}{\end{equation}}
\newcommand{\bea}{\begin{eqnarray}}
\newcommand{\eea}{\end{eqnarray}}

\newcommand{\baa}{\begin{array}}
\newcommand{\eaa}{\end{array}}
\long\def\symbolfootnote[#1]#2{\begingroup
\def\thefootnote{\fnsymbol{footnote}}\footnote[#1]{#2}\endgroup}
\newcommand{\laf}{\lambda_\phi}
\newcommand{\lam}{\lambda_m}
\newcommand{\las}{\lambda_\sigma}

\begin{document} 
\begin{flushright}
\end{flushright}

\bigskip\medskip

\thispagestyle{empty}

\vspace{2cm}

\begin{center}
  {\Large {\bf Standard Model  with spontaneously 

\bigskip
broken quantum scale invariance}}

\vspace{1.cm}

 {\bf D. M. Ghilencea}$^{\,a,b}$, \,\,  {\bf Z. Lalak}$^{\,c}$
  and \,\, {\bf  P. Olszewski}$^{\,c}$ 
\symbolfootnote[1]{E-mail: dumitru.ghilencea@cern.ch, 
zygmunt.lalak@fuw.edu.pl, pawel.olszewski@fuw.edu.pl}

\bigskip

$^a$ {\small  Theory Division, CERN, 1211 Geneva 23, Switzerland}

$^b$ {\small Theoretical Physics Department, National Institute of Physics}

{\small and Nuclear   Engineering  Bucharest\, 077125, Romania}

$^c$ {\small Institute of Theoretical Physics, Faculty of Physics, University of Warsaw}

{\small  ul. Pasteura 5, 02-093 Warsaw, Poland}

\end{center}

\begin{abstract}
\noindent
We explore the possibility that scale symmetry is a  quantum symmetry that is
broken {\it only} spontaneously and apply this idea to the Standard Model (SM).
We compute the quantum corrections to the potential of the higgs field ($\phi$)
in the classically scale invariant version of the SM 
($m_\phi=0$ at tree level)  extended by the dilaton ($\sigma$).
The tree-level potential of $\phi$ and $\sigma$, dictated by scale invariance,
may contain non-polynomial effective operators, e.g. $\phi^6/\sigma^2$, $\phi^8/\sigma^4$,
$\phi^{10}/\sigma^6$, etc.
The one-loop scalar potential is scale invariant, since the loop calculations
 manifestly preserve the scale symmetry, with the DR subtraction scale $\mu$
 generated spontaneously by the dilaton vev $\mu\sim\langle\sigma\rangle$.
The Callan-Symanzik equation of  the potential is  verified in the presence of the 
gauge, Yukawa and the non-polynomial operators. The couplings of the non-polynomial
 operators have non-zero beta functions that we can actually compute from the quantum potential.
At the quantum level the higgs mass  is protected by spontaneously broken
scale symmetry,  even though the theory is non-renormalizable.
 We compare the one-loop potential  to its counterpart computed in 
the ``traditional'' DR scheme that breaks 
scale  symmetry {\it explicitly} ($\mu=$constant)  in the presence at the tree level  of  the
non-polynomial operators.
\noindent
\end{abstract}

\newpage

\section{Motivation}

In this letter we explore the idea that scale symmetry is a quantum symmetry
and study its implications for physics beyond SM. However,  this symmetry
is broken in the real world.  We 
 shall consider here  only {\it spontaneous}  breaking of this (quantum)  symmetry\footnote{By 
quantum scale symmetry we mean that the full  1PI quantum action is scale invariant.}. 
One motivation of our study is that scale symmetry  plays a role in the
ultraviolet  (UV) behaviour of the models.
In particular the SM with a classical  higgs mass parameter $m_\phi\!=\!0$  has an 
increased symmetry: it is scale invariant at the tree level; this  was invoked \cite{Bardeen}
 to protect $m_\phi$ naturally \cite{hooft} from large quantum  corrections, but
 a full quantum study is needed.

Consider a classically scale invariant theory.
One known issue when studying scale symmetry at the  quantum level
is that the regularization of the loop corrections  introduces a dimensionful 
parameter  (subtraction scale $\mu$) which breaks {\it explicitly} the scale symmetry,
thus destroying the symmetry we want to investigate\footnote{One could use
a regularization that  does not keep manifest scale symmetry 
 and attempt to restore it ``by hand'' at the end, but
this  misses scale-invariant operators
if the theory is non-renormalizable (see later).}
and affecting the  UV properties of the quantum theory.
To avoid this, the UV regularization must  preserve this symmetry. 
This is done by  using  a subtraction  {\it function}  $\mu(\sigma)$
which generates   (dynamically) a  subtraction scale 
$\mu(\langle\sigma\rangle)$  when the field $\sigma$ acquires  a vev $\langle\sigma\rangle$
after spontaneous scale symmetry breaking. For details on this  see \cite{Englert}
and  recent examples at one-loop \cite{S1,S2,S3,tamarit,dg1} and higher loops \cite{Gretsch,dzp}.
Here $\sigma$ is the Goldstone mode (dilaton)
of the spontaneously broken scale symmetry\footnote{To be exact, the mass eigenstates 
may actually contain  a small mixing  of original $\phi$, $\sigma$.}.

The model we consider is a scale-invariant  SM, defined as SM with classical $m_\phi\!=\!0$
 and extended by the dilaton. The goal is to use this scale invariant regularization
to compute quantum corrections to the  scalar potential. The quantum
result is scale invariant, so it can only have {\it spontaneous} scale symmetry breaking, 
with  a flat direction for  the dilaton ($\sigma$).
For clarity, this result is then  compared to that in the ``usual''
dimensional regularization (DR) of $\mu$=constant scale,
which breaks  explicitly the scale symmetry at the quantum level.

Let us consider first a simplified scale invariant (classical)  
 theory (e.g. \cite{W}-\cite{Endo:2015ifa})  of two
 real scalar  fields $\phi$ (higgs-like) and $\sigma$.
The potential $V$ is an homogeneous function, having no dimensionful couplings, so
\bea\label{eq1}
V(\phi,\sigma)=\sigma^4\, W(\phi/\sigma),
\qquad\textrm{where}\qquad
W(\phi/\sigma)=V(\phi/\sigma,1)
\eea
We assume that  $V(\phi,\sigma)$ has spontaneous scale symmetry breaking
i.e.  that $\sigma$  acquires a non-zero vacuum expectation value
$\langle\sigma\rangle\not=0$.
We thus search for such a solution and for the necessary condition
for this spontaneous breaking to happen .
With $\langle\sigma\rangle\not=0$  it  is then easy to see that the minimum conditions
 $V_\sigma=V_\phi=0$ ($V_\alpha=\partial V/\partial\alpha$) are equivalent to
\bea\label{minEW}
W(\rho)=W'(\rho)=0, \qquad \rho\equiv \phi/\sigma.
\eea
These equations can have a common solution 
$\rho_0\!\equiv\!\langle\phi\rangle/\langle\sigma\rangle$,
if the couplings satisfy a particular condition (constraint), see below.
Then a {\it flat direction} exists in the plane $(\phi,\sigma)$ with $\phi~=~\rho_0\,\sigma$.
Indeed, if $(\langle\phi\rangle,\langle\sigma\rangle)$
is a ground state with  $V\!=\!0$ then so is 
 $( t \langle\phi\rangle, t \langle\sigma\rangle)$, $t$ real.
Also the second derivatives matrix $V_{\alpha\beta}$ wrt $\alpha,\beta=\phi,\sigma$ has
 $\det (V_{\alpha\beta})\propto (4 W W^{\prime\prime}-3 W^{\prime\,2})=0$ on the ground state, 
so a massless state is indeed 
present corresponding to the flat direction.
Finally, since $\rho_0$ is a root of both $W$ and of its derivative $W'$, then 
$W(\phi/\sigma)\propto (\phi/\sigma-\rho_0)^2$, while if $V$ depends only on even powers
of the scalar fields (our model below), then the general structure is
\bea\label{www}
W(\phi/\sigma)\propto (\phi^2/\sigma^2-\rho_0^2)^2.
\eea
Note that the vanishing vacuum energy
$V(\langle\phi\rangle,\langle\sigma\rangle)\!=\!0$ 
follows from the (spontaneously broken) scale symmetry, see eq.(\ref{minEW}).
A scale invariant regularization of this theory leads to a scale invariant
quantum potential, which thus remains of the form in eq.(\ref{eq1}). 
Hence the above discussion around eqs.(\ref{eq1}), (\ref{minEW}), (\ref{www})
{\it remains true  at the quantum level}, including the possibility of
spontaneous-only breaking of the scale symmetry.

One of the two minimum 
conditions  in (\ref{minEW}) fixes the ratio $\rho_0\!=\!\langle\phi\rangle/\langle\sigma\rangle$
in terms of the (dimensionless) couplings of the theory. Thus all vev's of such theory, 
including  $\langle\phi\rangle$ are proportional to  $\langle\sigma\rangle\!\not=\!0$
which is a (unknown)  parameter of  the theory.
The second  minimum condition, after eliminating $\rho_0$ between the two equations
 in (\ref{minEW}),  gives a relation among the  couplings of the theory 
in the order of perturbation  in which $V$ is computed. 
This means that one coupling, say $\lambda_\sigma$ (the dilaton self-coupling)
is defined in terms of the rest $\lambda_\sigma\!=\!f(\lambda_{j\not=\sigma})$.
This relation follows from  demanding that $V$ have a {\it flat direction}\footnote{
This agrees with the {\it classical} result of \cite{Fubini} 
that spontaneous breaking of conformal symmetry to Poincar\'e symmetry 
for a single scalar field happens if the dilaton potential 
 $g \chi^{2d/(d-2)}$ has $g=0$ (flat direction); (note our model
 is a non-renormalizable  scale invariant quantum theory (see later)
 rather than  conformal).} 
which is a consequence of our requiring that quantum  scale symmetry
be broken  spontaneously. 
Such relation can be ``arranged'' by one initial classical tuning, 
with subsequent (quantum) tunings bringing ``acceptable'' 
$\cO(\lambda_j)$ corrections to this relation,
relative to the previous perturbation order\footnote{Perturbativity 
$\lambda_\sigma\!<\!1$ is maintained
for a weak coupling $\lambda_m$ between the visible $(\phi$) and hidden ($\sigma$) sectors,
see  later,  eq.(\ref{cond}) (a) or eq.(\ref{min}) (a) which 
fixes $\lambda_\sigma\!<\!1$ in terms of the other couplings,
for a small enough $\lambda_m$. 
}; this tuning ensures a vanishing  vacuum energy
 $V(\langle\phi\rangle,\langle\sigma\rangle)\sim W(\rho_0)=0$,
see conditions (\ref{minEW}).

We  stress  that the above picture, that builds on previous studies
\cite{Englert,S1,S2,S3,tamarit,dg1,Gretsch,dzp},
is very different from that obtained in the ``traditional'' DR scheme 
($\mu=$constant scale)
that is often used in classically scale invariant models
 e.g.\cite{Foot:2007iy}-\cite{Endo:2015ifa}; 
in such models scale symmetry is broken  {\it explicitly} by the (regularization of) 
quantum effects and then conditions (\ref{eq1}), (\ref{minEW})  are not true anymore 
at quantum level and the flat direction
is lifted by quantum corrections\footnote{The dilaton is then a  pseudo-Goldstone 
which is light, so it is regarded as the SM Higgs in those models.}.

What about the hierarchy problem? In the absence of gravity (not included here),
the Standard Model  has no hierarchy problem. However, 
this situation is no longer true under the reasonable assumption that 
 there is some  ``new physics'' beyond SM, e.g. a large vev of
a new scalar field  that couples to Higgs, etc.
In the model we consider,  defined by
 the scale invariant version of the SM  extended by the dilaton,
 we  have  ``new physics''  beyond the SM, represented
by the vev  $\langle\sigma\rangle$ that 
breaks  spontaneously the  scale symmetry. $\langle\sigma\rangle$ 
can  be very large compared to $\langle\phi\rangle$ 
where\footnote{
In the Brans-Dicke-Jordan theory of gravity, not considered here, one expects
$\langle\sigma\rangle\sim M_{\rm Planck}$.} the latter fixes  the electroweak scale\footnote{
Such hierarchy  can be  generated dynamically  \cite{FHR,FHR1} or as in \cite{K}.};
($\langle\sigma\rangle$ can then be regarded as a physical cut-off of the theory).
We simply  enforce such hierarchy by  choosing a very weak coupling of the visible to the
hidden sector of the dilaton\footnote{One takes $\vert\lambda_m\vert\!\ll\!\lambda_\phi$;
$\lambda_m$: coupling of hidden ($\sigma$)  to visible sector ($\phi$);
 $\lambda_\phi$: higgs self-coupling, see later.
The hierarchy of vev's (scales) is then  replaced by a (more fundamental)
hierarchy of  dimensionless couplings.} \cite{GGR}.
Such hierarchy is however  stable under quantum corrections, so
$m_\phi\sim\langle\phi\rangle\ll \langle\sigma\rangle$ without 
tuning at the quantum level \cite{S1,dg1}  and we verify this in our model at one-loop. 
This is expected to remain true to all
orders in perturbation theory since scale symmetry is preserved by the regularization and
is broken only spontaneously\footnote{
Scale symmetry may also be broken at some high scale due to Landau poles
of some of the couplings of the theory or due to other non-perturbative effects. 
We do not  consider these effects here since they involve physics above Planck scale
 in which case the present flat  space-time picture is not appropriate  - one needs to 
upgrade this formalism to include Brans-Dicke-Jordan gravity, see e.g. \cite{FHR}.}.
We thus have an example of a quantum stable hierarchy, with  a vanishing
 vacuum energy  at the loop level, that follow from  
the demand of spontaneously broken quantum scale symmetry.

In the following we apply these ideas to the scale invariant version 
of the SM  (with classical higgs mass $m_\phi\!=\!0$) extended by the dilaton.
The higgs and the dilaton have a  potential  dictated solely
 by the classical scale symmetry, so it can contain 
higher dimensional {\it non-polynomial} operators such as
 $\phi^6/\sigma^2$,\, $\phi^8/\sigma^4$, etc.
We then compute the one-loop potential with  a scale invariant regularization, 
so a flat direction is maintained at the quantum level.  
Even if the tree-level potential does not include the 
non-polynomial operators (by tuning their couplings to 0), they are generated at one-loop  with 
finite coefficient  \cite{dg1}  or as two-loop or higher counterterms \cite{Gretsch,dzp} - this
means the scale invariant quantum theory is non-renormalizable.
Further, the  quantum consistency of the theory is shown by
verifying the Callan-Symanzik equation of the potential
 in the  presence of the non-polynomial effective operators,
 gauge and Yukawa interactions. 
We also  compare the scale-invariant one-loop potential 
 to its counterpart computed in the ``usual'' DR scheme  that breaks scale 
symmetry explicitly ($\mu$=constant), in the presence 
at tree level of these effective operators.

If scale symmetry is preserved by  one-loop $V$,
 there is no dilatation anomaly which is a result of {\it explicit} scale symmetry breaking
by quantum calculations with $\mu$=constant. Contrary to common lore, 
the couplings still run  with  momentum \cite{S3,tamarit,dg1} 
since the vanishing of the beta functions  {\it is not a  necessary} condition
 for scale invariance. Their  one-loop running
is identical to that   in the theory with explicit  scale symmetry breaking ($\mu$=constant),
but at two-loop they start to differ in theories with
spontaneous versus  explicit breaking~\cite{tamarit,dzp}.

This analysis in flat space-time should be extended to  include the effects of gravity which 
we ignored. Since Einstein gravity breaks scale symmetry, a
 natural setup to include such effects
 is  to consider the Brans-Dicke-Jordan theory of gravity, see examples in
 \cite{S1,FHR,FHR1,Kannike:2016wuy,S4,S7,S8,O1,O2,O3}.
In such setup it may still be possible to perform a
 scale-invariant regularization  and then examine such scale invariant theory
 at quantum level.

\section{SM with a scale invariant one-loop potential}

\subsection{The tree-level scale invariant  potential}

Consider the SM Lagrangian  with tree-level higgs mass  $m_\phi=0$, so it is scale invariant.
 The higgs sector is weakly coupled to the ``hidden'' sector of the dilaton $\sigma$ with
\bea
\cL=\vert D_\mu H\vert^2+\frac{1}{2}(\partial_\mu\sigma)^2-V_0
\eea
where
\bea
H=\left(
\begin{array}{c}
G^+\\
\frac{1}{\sqrt 2}(\phi + i\,G^0)
\end{array}
\right)
\eea
and
\bea
\label{pot0}
V_0 &=&
\frac{\lambda_\phi}{3!}\, (H^\dagger H)^2
+
\frac{\lambda_m}{2}\, (H^\dagger H)\, \sigma^2
+
\frac{\lambda_\sigma}{4!}\,\sigma^4
+\frac{4\,\lambda_6}{3}\,\frac{(H^\dagger H)^3}{\sigma^2}
+\cdots
\eea
where the dots stand for higher powers of $H^\dagger H$. 
The neutral higgs ($\phi$) and dilaton  part is
\bea\label{pots}
V(\phi,\sigma)
&=&
\frac{1}{4!}\,\lambda_\phi\,\phi^4+\frac{1}{4}\,\lambda_m\,\phi^2\,\sigma^2+\frac{1}{4!}\,\lambda_\sigma
\sigma^4+
\frac{\lambda_6}{6}\,\frac{\phi^6}{\sigma^2}+....
\eea
We take $\lambda_\phi, \lambda_\sigma>0$ and $\lambda_m<0$  
and  that the two sectors of
$\phi$ and $\sigma$  are weakly coupled, with $\vert\lambda_m\vert <\lambda_\phi$. 
Regarding the terms  suppressed by powers of  $\sigma$,
they respect the (classical) scale symmetry of the action, so
they can be present in  the theory. They are well-defined\footnote{
Even if we set $\lambda_{6,8,...}=0$ at  EW scale, such terms are generated in a 
quantum scale invariant theory at one-loop (with a finite coefficient) \cite{dg1}
 or as  two-loop counterterms \cite{dzp}, so their presence is inevitable.
If   instead $\mu=$constant (explicit breaking)
and $\lambda_{6,8,...}\!=\!0$, such terms are never generated at quantum level.}
since  $\sigma$ acquires spontaneously a  vev  $\langle\sigma\rangle\not=0$
under conditions that we identify shortly (see
$(a)$ in eqs.(\ref{cond}), (\ref{min}) below).
One can expand such terms about the ground state, into an infinite sum of 
familiar  polynomial  (effective) operators:
\medskip
\bea\label{fluc}
\lambda_6\frac{\phi^6}{\sigma^2}=\lambda_6\frac{\phi^6}{\langle\sigma\rangle^2} \Big(
1-2 \frac{\sigma'}{\langle\sigma\rangle}+
3\frac{\sigma^{\prime 2}}{\langle\sigma\rangle^2}+
\cdots\Big),\qquad \sigma=\langle\sigma\rangle+\sigma',
\,\,\,\,\,\sigma':\textrm{fluctuation.}
\eea

\medskip\noindent
However, we prefer to use the form in eq.(\ref{pots})
since it keeps manifest the scale symmetry  of $\cL$.
Finally, we keep $\lambda_6\!\not=\!0$
but set to 0 the coefficients of $(H^\dagger H)^4/\sigma^4$ and higher terms.

Consider first $\lambda_6=0$.
We demanded spontaneous breaking of scale symmetry, so we seek the condition for which
 $\langle\sigma\rangle\not=0$. The minimum of $V$ exists if derivatives  $V_\phi=V_\sigma=0$,
giving
\bea\label{cond}
(a):\,\,\,\, \lambda_\sigma=\frac{9\lambda_m^2}{\lambda_\phi}\,\Big[1+\text{loops}\Big]\qquad
\text{and}\qquad
(b):\,\,\,\,
\frac{\langle\phi\rangle^2}{\langle\sigma\rangle^2}
=\frac{-3\lambda_m}{\lambda_\phi}\big[ 1+\text{loops}\big],
\eea
so also $\langle\phi\rangle\not=0$; here ``loops'' stands for loop corrections.

Let us then  assume that  $\lambda_\sigma$ 
is indeed that of   $(a)$ up to\footnote{It is actually
the generalization of $(a)$ for $\lambda_6\not=0$ that we shall assume to be true, see later.}
``loop'' effects that one can identify order by order in perturbation theory and 
that we ignore for the classical discussion here.
If (a) is true, we have spontaneous breaking of scale symmetry and
\medskip
\bea
\label{pot}
V=\frac{1}{4!}\,\lambda_\phi \sigma^4 \,\Big(\frac{\phi^2}{\sigma^2}
+\frac{3 \lambda_m}{\lambda_\phi}\Big)^2
\eea

\medskip\noindent
with $V=0$ at the  minimum.
A flat direction, corresponding to the Goldstone of scale symmetry (dilaton)
 exists in the plane $(\phi,\sigma)$.
The neutral higgs acquires a mass 
$m_{\tilde \phi}^2=(\lambda_\phi/3) (1-3\lambda_m/\lambda_\phi)\langle\phi\rangle^2$,
while the EW Goldstone bosons are massless. 
Thus, spontaneous scale symmetry breaking
triggers EW symmetry breaking, with a vacuum energy $V=0$.

Consider now  $\lambda_6\!\not=\!0$, 
with $\lambda_6\!>\!0$  for a well-defined $V$  at large $\phi$.
Then eqs.(\ref{cond}) become\footnote{
Eqs.(\ref{min}) for small $\lambda_6$ become
$\lambda_\sigma =(9 \lambda_m^2/\lambda_\phi)\big[1+\cO(\lambda_6)+\text{loops}\big]$
and
$\rho_0^2 = (-3 \lambda_m/\lambda_\phi)\big[1+\cO(\lambda_6)+\text{loops}\big]$.}
\medskip
\bea\label{min}
(a):\,\,\,\,\lambda_\sigma&=& \rho_0^2 \,\Big[\, 2 \lambda_6 \, \rho_0^4 - 3 \,\lambda_m\Big]
+\text{loops},
\qquad \text{where}
\nonumber\\[-3pt]
(b):\,\,\,\,\rho_0^2&\equiv&\frac{\langle\phi\rangle^2}{\langle\sigma\rangle^2}=
\frac{1}{12\lambda_6} \Big[
-\lambda_\phi+(\lambda_\phi^2-72\lambda_6\,\lambda_m)^{1/2}
\Big]
+\text{loops}.
\eea

\medskip
We assume from now on that $\lambda_\sigma$ is indeed given by relation   $(a)$, up to
small quantum corrections (ignored here), to ensure spontaneous scale symmetry
breaking;  this relation is ``protected'' by   scale symmetry.  The potential is then
\bea
V=\frac{\lambda_6}{6}\sigma^4\,\Big(\frac{\phi^2}{\sigma^2}-\rho_0^2\Big)^2
\Big(\frac{\phi^2}{\sigma^2}+\xi_0^2\Big),
\eea

\medskip\noindent 
in agreement with (\ref{www}).
Here $\xi_0^2\!=\!\big(\lambda_\phi+2 \,(\lambda_\phi^2-72\lambda_6\lambda_m)^{1/2}\big)/(12\lambda_6)\!>\!0$. 
If $\lambda_6\!\ra\! 0$ one recovers eq.(\ref{pot})\footnote{
With $y=\phi^2/\sigma^2$,   $V=(\lambda_\phi\sigma^4/4!)\, (y+3\lambda_m/\lambda_\phi)
\big[y+3\lambda_m/\lambda_\phi
+(4\lambda_6/\lambda_\phi) (y^2- 3 y\lambda_m/\lambda_\phi+9\lambda_m^2/\lambda_\phi^2)\big]\!+\!\cO(\lambda_6)$.}.
The neutral higgs mass can again be computed and 
recovers the above value for small\footnote{
One has 
\vspace{-0.2cm}
\bea\label{eq9}
m_{\tilde\phi}^2=\Big(-2+\frac{\lambda_\phi}{6\lambda_6}\Big)
\lambda_m\langle\sigma\rangle^2+\rho_0^2\,\Big[\frac{\lambda_\phi}{3}
\Big(\frac{\lambda_\phi}{6\lambda_6}-1\Big)-2\lambda_m\Big]\langle\sigma\rangle^2
=-\lambda_m \, \Big(1-\frac{3\lambda_m}{\lambda_\phi}\Big)\langle\sigma\rangle^2
+\cO(\lambda_6).\eea
\vspace{-0.3cm}
} $\lambda_6$;
the dilaton is again massless, with the flat direction mildly
changed by $\lambda_6$.
To conclude, spontaneous scale symmetry breaking
triggers EW symmetry breaking and ensures $V\!=\!0$ on the ground state. We would like
to know if this can remain  true at  quantum level.

The scale $\langle\sigma\rangle$ of ``new physics'' beyond SM should
be  larger than $\langle\phi\rangle\sim\cO(100)$ GeV. In Brans-Dicke-Jordan theory
of gravity (not considered here) that can generalise this study,
 one actually expects  $\langle\sigma\rangle\sim M_\text{Planck}$.
So a hierarchy $\langle\phi\rangle\ll\langle\sigma\rangle$
 may be generated dynamically \cite{FHR,FHR1}. Here we  
take a common view of a  very weak coupling of the hidden ($\sigma$) to visible ($\phi$) sector:
$\vert\lambda_m\vert\ll\lambda_\phi$ \cite{GGR};
 then\footnote{This hierarchy is stable under 
renormalization group \cite{GGR} due to  a shift symmetry, $\sigma\ra \sigma+\text{constant}$.}
from eq.(\ref{min})  $\lambda_\sigma\ll\vert \lambda_m\vert$. 
This classical ``tuning''\footnote{Note this is not a tuning in the sense of cancellation
of mass scales, seen in the mass hierarchy problem.} can ensure a hierarchy of scales
$\langle\phi\rangle\ll\langle\sigma\rangle$
($\lambda_6$ only brings sub-leading corrections, since the 
hierarchy is controlled by $\lambda_m$, the main 
coupling of the two sectors).

This concludes the picture of the classical potential with scale symmetry. At the quantum level, 
one question is whether the (quantum) scale symmetry, when spontaneously broken,
 maintains  the  hierarchy
$m_{\tilde \phi}^2\sim \langle\phi\rangle^2\ll\langle\sigma\rangle^2$ 
  without  additional tuning of the couplings.
If quantum corrections $\lambda_\phi^2\langle\sigma\rangle^2$ are generated, 
a tuning of the higgs self-coupling $\lambda_\phi$ would be needed and 
this would re-introduce the hierarchy problem.

\vspace{1cm}

\subsection{The one-loop scale invariant potential}

Let us compute the one-loop potential by preserving scale symmetry at quantum level
and thus avoid its explicit breaking by the UV regularization.
The method is described in  \cite{S1,S3,tamarit,dg1,Gretsch,dzp}.
To do this note we already have a vev $\langle\sigma\rangle$ that can act as 
subtraction scale.
The starting point is in $d=4-2\epsilon$ dimensions where
 the tree level potential is modified into
\medskip
\bea\label{tV}
\tilde V= \mu(\sigma)^{2\epsilon}\,V,\qquad\qquad
 \mu(\sigma)=z\,\sigma^{1/(1-\epsilon)},
\eea

\medskip\noindent
 $\tilde V$ is thus scale invariant in $d=4-2\epsilon$.
 The function $\mu(\sigma)$  generates a subtraction scale
$\mu(\langle\sigma\rangle)$ when $\sigma$ acquires a vev spontaneously.  
The definition of $\mu(\sigma)$ follows on dimensional grounds, with $z$ an arbitrary
 dimensionless subtraction parameter \cite{tamarit}. If we set 
$\mu(\sigma)$=constant, 
one immediately recovers the 
``traditional'' DR scheme  that breaks explicitly  the
 scale symmetry  in $d=4-2\epsilon$. 
We thus have two possible analytical continuations to 
$d=4-2\epsilon$ of the classical scale invariant theory in $d=4$: one 
is scale invariant (eq.(\ref{tV})), the other is not ($\mu$=constant),
and they lead to  distinct quantum  theories 
(of different symmetry) \cite{dg1,dzp},  as discussed below. 
The one-loop potential in $d=4-2\epsilon$ is then \cite{dg1,dzp}
\footnote{ $V_1$ is derived in $d\!=\!4-2\epsilon$ via usual
diagrammatic or functional methods in effective theories and
 remains valid in the presence of $\lambda_6\phi^6/\sigma^2$ term
which  is just a sum of familiar polynomial operators, see eq.(\ref{fluc}).}
\medskip
\bea
V_1&=&\tilde V -
\frac{i}{2}\,
\,\int \frac{d^d p}{(2\pi)^d}
\,{\rm Tr}\ln \big[ p^2-\tilde V_{ij}+i\varepsilon\big].
\eea

\medskip\noindent
This is computed in the Landau gauge.
The field dependent squared masses are eigenvalues of the matrix of
second derivatives denoted\footnote{
$\tilde V_{ij}
=\mu^{2\epsilon}\,\big[V_{ij} + 2\epsilon \,\mu^{-2}\,N_{ij}\big]+\cO(\epsilon^2)$;
$
N_{ij}\equiv \mu\, \Big\{
\,\frac{\partial\mu}{\partial s_i}\, \frac{\partial V}{\partial s_j}
 +\frac{\partial\mu}{\partial s_j}\, \frac{\partial V}{\partial s_i}\Big\}
 +\Big\{\mu\, \frac{\partial^2 \mu}{\partial s_i\partial s_j}
  -\frac{\partial \mu}{\partial s_i}\frac{\partial \mu}{\partial s_j}
\,\Big\}V.$}
$\tilde V_{ij}$
where subscripts  $i,j$ stand for: the 
 EW Goldstone scalars $G^0$, $\Re(G^+)$, $\Im(G^+)$, neutral higgs $\phi$ 
and dilaton $\sigma$. Unlike
 the  EW Goldstone modes or fermions and gauge bosons, the field-dependent masses of
  $\phi$ and $\sigma$ acquire a correction $\propto\!\epsilon$ relative 
to their values induced by $V$ alone, from derivatives of $\mu(\sigma)$
\bea\label{ev}
m_t^2&=&\frac{\mu(\sigma)^{2\epsilon}}{2} \,h_t^2 \phi^2,\,\quad
m_W^2=\frac{\mu(\sigma)^{2\epsilon}}{4}\,  g^2_2\,\phi^2,\,\quad
m_Z^2=\frac{\mu(\sigma)^{2\epsilon}}{4} \, (g_1^2+g^2_2)\,\phi^2,\,\nonumber\\
m_G^2&=&
\frac{\mu(\sigma)^{2\epsilon}}{6}\,\Big[\lambda_\phi\,\phi^2+3\lambda_m\sigma^2+
6\,\lambda_6\,\frac{\phi^4}{\sigma^2}\Big],
\nonumber\\
M^2_k&=&\mu(\sigma)^{2\epsilon}\,\Big[ m_k^2 +  \epsilon\, \delta_k\Big], \qquad k=\phi, \sigma.
\eea   

\medskip\noindent
where $m_t$ ($h_t$) is the field-dependent
 top mass (Yukawa coupling), $m_{W,Z}$ denote the gauge boson masses and 
$m_G$ denote the three  EW Goldstone field-dependent masses. Finally
$M_k^2$ are eigenvalues of   $\tilde V_{\alpha\beta}$, while
$m_k^2$ are  eigenvalues of the $2\times 2$ sub-matrix 
$V_{\alpha\beta}$ of $V_{ij}$ 
with\footnote{In general, in terms of derivatives of 
tree level V:
$m_k^2=\frac{1}{2} \big[ \Tr (V_{\alpha\beta}) \pm  \big[ (\Tr V_{\alpha\beta})^2
-4 \det V_{\alpha\beta}\big]^{1/2}\big]$
and also 
$\delta_k=
\mu(\sigma)^{-2}
\big\{
\Tr (N_{\alpha\beta}) \pm [(\Tr V_{\alpha\beta})( \Tr N_{\alpha\beta})-2\rho]/[(\Tr V_{\alpha\beta})^2
-4\det V_{\alpha\beta}]^{1/2}\big\}$.
The expression of $\rho$ is
$\rho=V_{\phi\phi} N_{\sigma\sigma}+ V_{\sigma\sigma}\,N_{\phi\phi}-2 V_{\phi\sigma} N_{\phi\sigma}
$, where  $N_{\phi\phi}=0$, $N_{\sigma\sigma}=z^2\,(2\sigma\, V_\sigma -V)$, 
$N_{\phi\sigma}=z^2\,\sigma V_\phi$.}
$V_{\alpha\beta}=\partial^2  V/\partial \alpha\, \partial \beta$,
$\alpha,\beta=\phi,\sigma$.
Then, one finds  at one-loop ($\kappa=(4\pi)^2$)
\medskip
\bea\label{lt}
V_1 
=\mu(\sigma)^{2\epsilon}
\Big\{
V- \frac{1}{4 \kappa}\,
\Big[
 \sum_{j=\phi,\sigma; G,W,Z,t} \!\!\!
n_j \, m^4_j(\phi,\sigma) \,  \Big(\,  \frac{1}{\epsilon}
- \ln\frac{m_j^2(\phi,\sigma)}{c_j\,\mu^2(\sigma)} \Big)
+
\frac{4 \,(V_{\alpha\beta}\,N_{\beta\alpha})}{\mu^2(\sigma)}\Big]
\Big\}.
\eea

\medskip\noindent
 with summation over $\alpha,\beta=\phi,\sigma$ and $N_{\alpha\beta}=
\mu (\mu_\alpha V_\beta+\mu_\beta V_\alpha)-\mu_\alpha\mu_\beta)\, V$ and
$\mu_\alpha=\partial\mu/\partial \alpha$. 
Also  $n_j=\{3,1,6,3,-12\}$ for $j=\{G,S,W,Z,t\}$, with $S=\phi,\sigma$;
$c_j=4\pi e^{3/2-\gamma_E}$ if $j=\phi,\sigma, t,G$ and $c_j=4\pi e^{5/6-\gamma_E}$ if  $j=W,Z$.
The one-loop term $(V_{\alpha\beta}\,N_{\beta\alpha})$ is a new correction,
 absent in the case of $\mu$=constant (i.e. explicit scale symmetry breaking by the regularization).

The poles in the one-loop Lagrangian are cancelled
by  the counterterm  $\delta L_1$\footnote{One can  use
 $\sum_{k=\phi,\sigma} m_k^4=V_{\phi\phi}^2+V_{\sigma\sigma}^2+2 V_{\phi\sigma}^2$.}
\medskip
\bea\label{ZV}
\delta L_1 & \equiv& 
\frac{1}{2} (Z_\phi-1)\,(\partial_\mu\phi)^2
+\frac{1}{2} (Z_\sigma-1)\,(\partial_\mu\phi)^2
\\[-3pt]
&-& \mu(\sigma)^{2\epsilon} 
\Big\{\, \frac{1}{4!} (Z_{\lambda_\phi}\!-1) \lambda_\phi\phi^4+
 \frac{1}{4} (Z_{\lambda_m}\!-1)\lambda_m \phi^2\sigma^2
+\frac{1}{4!} (Z_{\lambda_\sigma}\!-1)\lambda_\sigma\sigma^4\big)
 \nonumber\\[-3pt]
&+&
\sum_{j=3,4,5,6} 
\frac{1}{2 j}\, (Z_{\lambda_{2 j}}-1)\, \lambda_{2j}\,\, \frac{\phi^{2j}}{\sigma^{2j-4}}
\Big\}.\nonumber
\eea 

\bigskip\noindent
Introducing the notation:
\bea\label{gZ}
Z_\xi=1+\frac{1}{\epsilon}\,\frac{\gamma_\xi}{\kappa},\qquad \xi=\lambda_\phi,\lambda_\phi, \,\text{etc}.
\eea
one identifies:
\medskip
\bea
\gamma_{\lambda_\phi}&=&\frac{3}{2 \lambda_\phi}
\,\Big(\,
\frac{3}{2} g_2^4
+\frac{3}{4}\, (g_1^2+g_2^2)^2-12 h_t^4+\frac{4}{3}\lambda_\phi^2
+\lambda_m^2
+32 \lambda_m\lambda_6 
\Big),
\nonumber\\[-3pt]
\gamma_{\lambda_m}&=&
\frac{1}{2}\,(2\lambda_\phi+\lambda_\sigma+ 4\lambda_m),
\nonumber\\[-3pt]
\gamma_{\lambda_\sigma}&=&\frac{3}{2}
\,(\lambda_\sigma+4 \lambda_m^2/\lambda_\sigma).
\eea

\medskip\noindent
Notice that $\lambda_6$ contributes  to 
$\gamma_{\lambda_\phi}$ and  to the beta function of $\lambda_\phi$ (see later). Finally
\medskip
\bea
 \gamma_{\lambda_6}&=& \frac{3}{2}
\,( 6\lambda_\phi-8\lambda_m+\lambda_\sigma),\hspace{6.3cm}
\nonumber\\[-3pt]
 \gamma_{\lambda_8}&=&\frac{2 \,\lambda_6}{\lambda_8}\, (28 \lambda_6+\lambda_m),
\nonumber\\[-3pt]
\gamma_{\lambda_{10}}&=& 20 \frac{\lambda_6^2}{\lambda_{10}}, 
 \qquad
\gamma_{\lambda_{12}}\,= \,\frac{3 \lambda_6^2}{\lambda_{12}}.
\label{Z}
\eea

\medskip\noindent
Therefore, the non-polynomial operator  $\lambda_6\phi^6/\sigma^2$ in the tree-level $V$ generated 
new non-polynomial counterterms up to and including $\phi^{12}/\sigma^8$,
 of couplings $\propto\!\lambda_6$. 
This effect is independent
of whether the quantum calculation respects or not  the scale symmetry (i.e. $\mu\!\sim\!\sigma$
or $\mu$=constant). The generalisation to 
more such operators at the tree level is immediate.

The SM one-loop  potential $U_1$  is then
\medskip
\bea\label{U}
U_1&= &V+V^{(1)}+V^{(1,n)},
\eea
where
\bea
V^{(1)} &= & \frac{1}{4\kappa}\,
\sum_{j=\phi,\sigma; G,t,W,Z} n_j\, m^4_j(\phi,\sigma)\, \ln \frac{m^2_j(\phi,\sigma)}{c_j\,(z\,\sigma)^2},
\\
V^{(1,n)}
\!\!\!& =&\!\!\!   \frac{1}{48 \kappa} 
\Big[
(-16 \lam \laf- 18 \lam^2+\laf \las)\,\phi^4
-
\lam (48 \lam+25 \las) \phi^2\sigma^2
-
7 \las^2\sigma^4
\\
&+&
(\laf \lam +6 \lambda_6\las)\frac{\phi^6}{\sigma^2}
+
8\,\lambda_6\,(4\laf -2 \lam)\frac{\phi^8}{\sigma^4}
+
\lambda_6\,(192 \lambda_6+2\lambda_\phi)\frac{\phi^{10}}{\sigma^6}
+
40\lambda_6^2\frac{\phi^{12}}{\sigma^8}
\Big].\nonumber
\eea

\medskip\noindent
$U_1$ is manifestly scale invariant.
Firstly, the Coleman-Weinberg (CW) term is  modified into a scale invariant form $V^{(1)}$
where we finally  replaced  $\mu(\sigma)=z\,\sigma$ 
(see (\ref{tV}) for $\epsilon\ra 0$). Note that $V^{(1)}$  contains new terms of
the form $\phi^8/\sigma^4\ln[...]$, $\phi^6/\sigma^2\ln[...]$ of coefficients $\propto\lambda_6$,
that originate from $m_G^4(\phi,\sigma)$. 
In the ``usual'' DR scheme $V^{(1)}$ has the same form, with $(z\sigma)\ra \mu$.

There is also a finite  one-loop contribution $V^{(1,n)}$
 due to ``evanescent'' corrections ($\propto~\epsilon$) to 
the field-dependent masses of $\phi$ and $\sigma$  (eq.(\ref{ev})), induced by
 {\it derivatives} of $\mu\sim \sigma$.
Therefore, $V^{(1,n)}$ is not present in the other case of $\mu$=constant when
 the regularization  breaks the scale symmetry;  thus $V^{(1,n)}$ can distinguish between these
two cases at one-loop\footnote{These two cases are
 different quantum theories (have different symmetry).}. 
 Further, in the classical decoupling limit
of the hidden sector from the SM, $\lambda_m\!\ra\! 0$ and $\lambda_6\!\ra\! 0$,
then  $V^{(1,n)}$ vanishes.
$V^{(1,n)}$ also contains terms non-polynomial in fields 
like
$\lambda_m\lambda_\phi\phi^6/\sigma^2$ that remains present even if we set
$\lambda_6\!=\!0$\footnote{Assuming one set $\lambda_6\!=\!0$ at tree level,
some other subtraction scheme could eventually
  remove finite $\phi^4$, $\phi^2\sigma^2$ or $\sigma^4$ terms
in $V^{(1,n)}$, but could not remove the remaining $\lambda_m\lambda_\phi \phi^6/\sigma^2$ 
 that does not vanish for  $\lambda_6=0$.}. 
At two-loop such non-polynomial operators, including higher order $\phi^8/\sigma^4,$
etc, emerge as two-loop  counterterms \cite{dzp} even if we set
$\lambda_6\!=\!0$\footnote{
The two-loop beta functions of such terms are non-zero even if $\lambda_6=0$, so
 setting these to zero (at some scale)  will not remove them since  
they are again generated at a   different scale \cite{dzp}. }.

 Although we do not show it, one can immediately Taylor expand both $V^{(1)}$ and $V^{(1,n)}$ 
about the non-zero vev of $\sigma$, with
 $\sigma\!=\!\langle\sigma\rangle+\sigma'$ (and eventually of $\phi$ too,
$\phi=\langle\phi\rangle+\phi'$). 
  One then obtains a representation
that contains an infinite sum of {\it polynomial} operators in the field fluctuations
 ($\phi',\sigma'$) suppressed by powers of $\langle\sigma\rangle$. 
However, in this case manifest scale symmetry of the
quantum result is lost.

\subsection{One-loop beta functions and Callan-Symanzik equation}

To check the quantum consistency of the  scalar potential, we
 verify the Callan-Symanzik equation for it. This is to ensure
that physics is independent 
of the subtraction scale $\mu(\langle\sigma\rangle)=z\,\langle\sigma\rangle$.
To this purpose we need the one-loop beta functions of all couplings, including
those of the non-polynomial operators.
These are computed from the condition that the 
bare coupling is independent of the  subtraction parameter\footnote{
The dimensionless parameter $z$ tracks
the dependence on the subtraction scale $\mu(\langle\sigma\rangle)=z\langle\sigma\rangle$.
} $z$. For example
$d/(d\ln z) \lambda_\phi^B=0,$ where 
$\lambda_\phi^B=\mu(\sigma)^{2\epsilon} \lambda_\phi Z_\phi^{-2} Z_{\lambda_\phi}$
and $\phi_B^2=Z_\phi\phi^2$. Using these relations,  the beta function that is
$\beta_{\lambda_\phi}=d\lambda_\phi/d(\ln z)$ becomes
\bea\label{hh}
\beta_{\lambda_\phi}
=-2\epsilon\lambda_\phi+\frac{ 2\lambda_\phi}{\kappa}\,\alpha_j\,\frac{d}{d\alpha_j} 
\Big[ \gamma_{\lambda_\phi}-2 \gamma_\phi\Big],
\eea

\medskip\noindent
with summation over $j$ with $\alpha_j=g_1^2$, $g_2^2$, $h_t^2$, 
$\lambda_\phi,$ $\lambda_m,$ $\lambda_\sigma,\lambda_6,\lambda_8,$ etc. Next, using
notation (\ref{gZ}), one has
\bea\label{gammaphi}
\gamma_\phi=\frac{1}{\kappa} \Big(\frac{3}{4} g_1^2+\frac{9}{4} g_2^2- 3 h_t^2\Big),
\qquad
\gamma_\sigma=0
\eea

\medskip\noindent
which can easily be computed in a scale invariant way\footnote{$\gamma_\phi$ and $\gamma_\sigma$
have the same expression as when $\mu=$constant scale.}. 
Relations similar to eq.(\ref{hh}) exist for the other beta functions. We then find
\medskip
\bea\label{betas1}
\beta_{\lambda_\phi}&=&
\frac{2\lambda_\phi}{\kappa}\,
(\gamma_{\lambda_\phi} -2 \gamma_\phi),
\nonumber\\[-3pt]
\beta_{\lambda_m}&=&\frac{2\lambda_m}{\kappa}
 (\gamma_{\lambda_m}-\gamma_\phi),
\nonumber\\[-3pt]
\beta_{\lambda_\sigma}&=&
\frac{2\lambda_\sigma}{\kappa} \gamma_{\lambda_\sigma}.
\eea

\medskip\noindent 
$\beta_{\lambda_\phi}$ includes a correction due
to  $\lambda_6$, which is the coupling of the
non-polynomial term that we included in the classical potential eq.(\ref{pot0}).
These  one-loop beta functions are identical  to those of 
the similar theory with a regularization that breaks scale 
symmetry explicitly ($\mu=$constant)\footnote{
However,  at two-loop order the beta functions start to differ \cite{dzp}
in our case of spontaneous scale symmetry breaking from the case of explicit breaking 
(by the regularization with $\mu=$constant). In this case
the evanescent corrections $\propto\epsilon$
to scalar field-dependent masses (higgs, dilaton) of the potential
 ``meet'' the $1/\epsilon^2$ usual two-loop poles, to
bring  new poles $\epsilon\times 1/\epsilon^2\sim 1/\epsilon$ that demand new
counterterms, thus  modifying the beta functions, see \cite{dzp} for details.}.
We find in a similar way
\medskip
\bea\label{betas2}
\beta_{\lambda_6}&=&
\frac{2\lambda_6}{\kappa}\,(\gamma_{\lambda_6}-3 \gamma_\phi),
\nonumber\\[-3pt]
\beta_{\lambda_8} &=&
\frac{2\,\lambda_8}{\kappa}\,(\gamma_{\lambda_8}-4 \gamma_\phi),
\nonumber\\[-3pt]
\beta_{\lambda_{10}}&=&
\frac{2\lambda_{10}}{\kappa}(\gamma_{\lambda_{10}}-5 \gamma_\phi),
\nonumber\\[-3pt]
\beta_{\lambda_{12}}&=&
\frac{2\lambda_{12}}{\kappa}(\gamma_{\lambda_{12}}-6\gamma_\phi).
\eea

\medskip\noindent
These beta functions of the non-polynomial operators are
difficult to obtain by other methods (diagrammatic, etc). This justifies 
keeping these operators in a scale symmetric form (eq.(\ref{pots})), rather than expanding them
about the ground state in series of polynomial operators (eq.(\ref{fluc})).
 
The Callan-Symanzik equation of the scalar potential states the  independence
of the potential of the subtraction scale. At one-loop this gives \cite{tamarit}
\medskip
\bea
\frac{d\,}{d\ln z} U_1 (\phi,\sigma,\alpha_k)=
\Big( z\frac{\partial }{\partial z}+\beta_{\alpha_k}\frac{\partial}{\partial \alpha_k}
+\gamma_\phi\,\phi\,\frac{\partial}{\partial\phi}
\Big)\,
U_1(\phi,\sigma,\alpha_k)=\cO(\alpha_j^3).
\eea

\medskip\noindent
Here $\alpha_k$ denote the couplings $\lambda_\phi$, $\lambda_\sigma$, $\lambda_m$, $g_1^2$, $g_2^2$, $h_t^2$,
$\lambda_6$,...$\lambda_{12}$ which were found to have nonzero beta functions. Further
$\gamma_\phi=\frac{\partial \ln\phi}{\partial \ln z}=-\frac{1}{2}\frac{\partial 
\ln Z_\phi}{\partial\ln z}$ was found in eqs.(\ref{gammaphi}), (\ref{gZ}),
while $\gamma_\sigma=0$. Finally $U_1(\phi,\sigma,\alpha_k)$ denotes the potential found in eq.(\ref{U})
with the observation that all couplings are now replaced by 
their ``running'' versions. In particular the tree level potential (part of  $U_1$) is supplemented with
the following terms with running couplings $\lambda_{8,10,12}$
\bea
V\ra V+\frac{\lambda_8}{8}\frac{\phi^8}{\sigma^2}
+\frac{\lambda_{10}}{10}\frac{\phi^{10}}{\sigma^6}
+\frac{\lambda_{12}}{12}\frac{\phi^{12}}{\sigma^8}.
\eea

\medskip\noindent
These terms are present since the  couplings $\lambda_{8,10,12}$ (which had boundary values
set to $0$ at the EW scale, unlike $\lambda_6\not=0$),  have non-zero beta
functions.

The only explicit $z$-dependent part in $U_1$ comes through the
Coleman-Weinberg part $V^{(1)}$ of eq.(\ref{U}), while the terms involving the beta functions
 and anomalous dimension act only on the tree level part of the potential, 
in our one-loop approximation.

With the above results, checking the Callan Symanzik equation is  immediate.
We stress that this is verified in the presence of the non-polynomial 
operators that actually correspond to infinitely many polynomial operators when 
expanded about the ground state.

In conclusion,  the potential is indeed  independent  of the subtraction scale
$z\langle\sigma\rangle$, so one can take any value for it. It is customary 
 to set the subtraction scale  equal to $\langle\phi\rangle$, to minimise 
 the log terms in  the potential.
 In our scale invariant approach  $\mu(\sigma)=z\, \sigma$, so
after scale symmetry breaking $\mu(\langle\sigma\rangle)=\langle\phi\rangle$
if we take $z=\langle\phi\rangle/\langle\sigma\rangle$, and we  do so below.
This means the couplings and fields are evaluated at the scale $\langle\phi\rangle$.

\subsection{The one-loop higgs mass}

The one-loop corrected potential  is scale invariant and it
 has a flat direction\footnote{See previous discussion in the Introduction 
around  eqs.(\ref{minEW}).}, the dilaton, which remains
 massless  at the quantum level\footnote{This is not the case in 
the ``traditional'' DR scheme where scale symmetry is broken explicitly by the
quantum calculation with $\mu$=constant and 
the dilaton becomes a  pseudo-Goldstone mode.}.
We can compute the higgs mass $m_{\tilde\phi}$ at one-loop by using
\bea
m_{\tilde\phi}^2=(U_1)_{\sigma\sigma}+(U_1)_{\phi\phi} \Big\vert_{\rm min}
\eea

\medskip\noindent
where the subscripts denote derivatives with respect to the fields shown.
We calculate the new ground state and the correction  $\delta m_{\tilde \phi}^2$ 
to classical $m_{\tilde\phi}^2$
 in the limit of an ultraweak coupling of the visible to the hidden sector
$\vert\lambda_m\vert\ll \lambda_\phi$ (giving $\lambda_\sigma\ll\vert\lambda_m\vert$). This was
 motivated earlier in that it ensures a classical
hierarchy $\langle\phi\rangle\ll\langle\sigma\rangle$.
The new ground state is modified to
\medskip
\bea
\frac{\langle\phi\rangle^2}{\langle\sigma\rangle^2}=\frac{-3\lambda_m}{\lambda_\phi} 
\big[1+\zeta\big],\qquad
\zeta=-\frac{\lambda_\phi}{4 \kappa}[4\ln (\lambda_\phi/2)-8]
\eea

\medskip\noindent
With the notation $g^2\equiv g_1^2+g_2^2$, the one-loop correction  is
\medskip
\bea
\delta m_{\tilde\phi}^2&=&
\frac{-\lambda_m}{\lambda_\phi}\frac{\langle\sigma\rangle^2}{16\kappa}
\, \Big\{
\,27\, \Big[
g^4 \Big( \ln\frac{g^2}{4} +\frac{1}{3}\Big)
+2 \,g_2^4\, \Big(\ln \frac{g_2^2}{4}+\frac{1}{3}\Big)
-16 \,h_t^4\, \Big(\ln\frac{h_t^2}{2}-\frac{1}{3}\Big)\Big]
\nonumber\\[-3pt]
&&\qquad\qquad\quad +\,\,
4\,\lambda_\phi^2\, \Big[ 5 \ln \frac{\lambda_\phi^2}{12} - 8 +\ln 27\Big]
\Big\}.
\eea

\medskip\noindent
This quantum correction remains proportional to 
$\lambda_m\langle\sigma\rangle^2\sim \cO(100\text{GeV})$, just like the tree-level value.
Thus the initial classical hierarchy $m_{\tilde\phi}
\sim\langle\phi\rangle\ll \langle\sigma\rangle$
is stable in the presence of quantum corrections, without any quantum tuning  
of the couplings $\lambda_{\phi,m,\sigma}$, in agreement with previous
results \cite{S1,dg1}\footnote{
In particular there is no term
 $\lambda_\phi\langle\sigma\rangle^2$ that would require tuning the higgs self-coupling
$\lambda_\phi$, etc.}'\footnote{
For the physical higgs mass there is also the usual correction of running from 
$p^2=0$ to $p^2=m_h^2$.}.
An additional correction  from $\lambda_6\not=0$ does not change this result since
it is sub-leading
in the limit of ultraweak coupling considered here (being suppressed by the
large $\langle\sigma\rangle$). Finally, the spontaneous breaking of scale symmetry
used here avoids the constraint of \cite{Schmaltz} (derived using
explicit breaking by the ``usual'' DR scheme) that demands new physics 
at the TeV scale.

\subsection{What about the dilatation anomaly?}

Let us analyze  the situation of the dilatation current $D^\mu$
and its divergence \cite{S3,tamarit}.
 For a set of fields $\phi_j$ ($\phi$, $\sigma$, etc)
\bea
D^\mu& =& \frac{\partial \cL}{\partial (\partial_\mu\phi_j)}\,(x^\nu\partial_\nu\phi_j+d_\phi)-x^\mu\,\cL,
\nonumber\\
\partial_\mu D^\mu & = & 
(d_\phi+1)\,(\partial_\mu\phi_j)\,
\frac{\partial\cL}{\partial (\partial_\mu\phi_j)}+d_\phi\, \phi_j\,
\frac{\partial \cL}{\partial \phi_j}-d\cL,
\eea

\medskip\noindent
with $d_\phi$ the mass dimension of $\phi$, $d_\phi=(d-2)/2$ for a scalar in $d$ dimensions.
For standard kinetic terms 
and using the equations of motion, 
we find for  a potential $\cV$ in $d$ dimensions
\medskip
\bea\label{div}
\partial_\mu D^\mu= d \, \cV- \frac{d-2}{2} \phi_j\,\frac{\partial \cV}{\partial\phi_j}.
\eea

\medskip\noindent
Consider now that  $\cV$ is scale invariant 
at both classical and quantum level as in our case\footnote{
We have $\cV=\mu(\sigma)^{2\epsilon} \, V$  in eqs.(\ref{tV}), while in eq.(\ref{U}) 
 $\cV=\mu(\sigma)^{2\epsilon} U_1$  before  $\epsilon\ra 0$. },  eq.(\ref{tV}) (also
eq.(\ref{U})).
Therefore, for a  dimensionless  parameter $\rho$, $\cV$ has the property\footnote{This property
is shown using that: $\cV(\phi_j)=\phi_k^\xi\, \cV(\phi_j/\phi_k)$, $k=fixed$; since $[\cV(\phi_j)]=d$,
$\cV(\phi_j/\phi_k]=0$ and $[\phi_j]=(d-2)/2$ then $\xi=2d/(d-2)$.
Then $\cV(\rho \,\phi_j)=(\rho\,\phi_k)^\xi \cV(\phi_j/\phi_k)=
\rho^\xi\,\cV(\phi_j)$ with $\xi$ as above.} 
 $\cV(\rho\,\phi_j)=\rho^{2 d/(d-2)}\,\cV(\phi_j)$ 
 in $d=4-2\epsilon$ dimensions (homogeneous function).
Differentiating this equation  with respect to $\rho$ and then taking $\rho\ra 1$ gives
 $2 d/(d-2)\,\cV=\phi_j\,\partial \cV/\partial\phi_j$ so the
 rhs of eq.(\ref{div}) then  vanishes.
 Therefore  $\partial_\mu D^\mu=0$ at both classical and quantum level,
so there is no anomalous breaking of the  quantum scale symmetry. 
Nevertheless the couplings still ``run'' and  have non-zero beta functions 
(eq.(\ref{betas1})) with their corresponding poles in $\cL$.

To understand this better, let us also see  what happens if $\cV$ is
not scale invariant in $d=4-2\epsilon$. 
This happens  when  $\cV=\mu^{2\epsilon} V(\phi_j)$ which is the case of
  the ``traditional'' DR scheme with explicit scale symmetry breaking, 
with $\mu$   a  fixed scale (not a function of the fields)
and $V$ the  potential, scale invariant in $d=4$ (assuming no mass terms).
Then
$V(\rho\,\phi_j)=\rho^4\,V(\phi_j)$, but $\cV$ is no longer scale invariant in  
$d=4-2\epsilon$. Then from eq.(\ref{div})
\medskip
\bea
\partial_\mu D^\mu=d\,\mu^{2\epsilon} \,V
-2 (d-2)\,\mu^{2\epsilon} \, V(\phi_j)
=2\epsilon \,\mu^{2\epsilon}\, V=2\epsilon\,\mu^{2\epsilon} \lambda_j\frac{\partial V}{\partial\lambda_j}.
\eea
While at the classical level the rhs vanishes when $\epsilon\ra 0$,
 at the quantum level the quartic couplings $\lambda_j$ in $V$ acquire a pole
$\beta_{\lambda_j}/\epsilon$ which cancels the $\epsilon$ in front, to give a finite non-zero
 rhs $\partial_\mu D^\mu\propto\beta_{\lambda_j}(\partial\cL/\partial \lambda_j)$. This is the familiar
scale anomaly breaking of  the conservation of this current 
 in the ``traditional'' DR scheme
\footnote{even if at classical level it was conserved}$^,$\footnote{If $V$
contained mass terms,  $\partial_\mu D^\mu$ also
contains a ``classical'' breaking of scale symmetry term, $m^2\phi^2$.}.

In conclusion, it is   scale invariance of the action 
in $d\!=\!4-2\epsilon$ that ensures that no scale anomaly is
present. This invariance in $d\!=\!4-2\epsilon$ is lost in the ``usual'' DR
regularization with explicit breaking ($\mu=$constant).
Thus, the   vanishing of the beta function is not a  necessary condition
for the theory to be  scale invariant; one must also specify how the quantum  theory 
was regularized, with or without respecting its scale symmetry. 
In other words the non-vanishing of the beta function does not mean the theory cannot 
be scale invariant.

\subsection{Further remarks}

 As mentioned, the vacuum energy  vanishes in models
with scale symmetry or with spontaneous breaking of it, see discussion 
after eq.(\ref{minEW}).  This protection remains
in place at the quantum level provided this symmetry is respected
by the quantum calculation itself. The initial classical
 tuning of the boundary values of 
the  couplings, relation (a) in eq.(\ref{cond}):
 $\lambda_\sigma=9\lambda_m^2/\lambda_\phi(1+\cO(\lambda_6))$,
assumed to be true in the paper (for spontaneous breaking to exist),
receives  loop corrections of order $\cO(\lambda_j)$.
As a result the tuning of the couplings, that enforces $V_\text{min}=0$
at the loop level (demanded by scale symmetry), is $\cO(\lambda_j)$ relative to its tree level
case. More generally, in order $n$  the tuning is $\cO(\lambda_j)$ 
relative to that in order $n-1$, i.e. at the level of 
 the precision of the perturbation theory calculation in that order.

The consistency of the 
 boundary values for the running  couplings with some high scale 
physics  that must fix the value of $\langle\sigma\rangle$ 
should be investigated. This  discussion requires one extend this quantum
calculation to the case of curved space-time while respecting this symmetry.
 The appropriate setup is in the context of Brans-Dicke-Jordan theory of gravity. 
As discussed in \cite{FHR2}, in such frame with non-minimal couplings,
the dilaton (with derivative couplings) decouples and avoids ``fifth force experiments''.
For investigations along this direction see \cite{FHR,FHR1,
Kannike:2016wuy,S4,S7,S8,O1,O2,O3}\footnote{
Another possibility is to consider Einstein gravity which breaks
the scale symmetry discussed here. 
Then scale symmetry is only  an approximate symmetry 
 and the dilaton is a pseudo-Goldstone mode that acquires a small mass
and the vacuum energy is then non-zero.}.

\section{Conclusion}

We explored the possibility that scale symmetry is a quantum symmetry of the SM that
is broken (only) spontaneously. Following previous developments on this idea,
 we considered the case of the classically scale invariant version of the SM which has
vanishing tree-level mass for the higgs ($\phi$)  and  is extended
 by the dilaton $\sigma$ (the Goldstone mode
of scale symmetry). The vev $\langle\sigma\rangle\not=0$
breaks the  scale symmetry spontaneously and generates dynamically a subtraction 
scale $\mu\sim\langle\sigma\rangle$ that is necessary for quantum calculations.

The classical scalar potential is dictated by the scale symmetry only and may
contain non-polynomial effective operators such as $\lambda_6 \phi^6/\sigma^2$, 
$\lambda_8 \phi^8/\sigma^4$, $\lambda_{10}\phi^{10}/\sigma^6$, $\lambda_{12}\phi^{12}/\sigma^8$, etc;
these may always be Taylor-expanded into a sum of  infinitely-many 
polynomial  operators  in fields fluctuations
suppressed by powers of $\langle\sigma\rangle$ (which 
 can be regarded as  a physical cutoff of the theory);
however, in such case the manifest scale symmetry  of the theory is lost.

The one-loop computation of the potential respected the 
scale symmetry of the classical Lagrangian.
As a result, a  scale invariant   one-loop  
potential for the higgs and dilaton is obtained.
The quantum potential has corrections from gauge and Yukawa 
interactions and also from the higher dimensional,
  non-polynomial operators. 
The latter were included in the classical Lagrangian and
their couplings ($\lambda_6, \lambda_8, \lambda_{10}, \lambda_{12}$, etc) are one-loop renormalized with 
 beta functions that we computed from the quantum potential. These beta functions are  
difficult to compute by  other means and are an important result of this work.
Tuning these couplings  to zero at the tree-level will not avoid
 the  presence of their  corresponding operators at the quantum level;
 these operators re-emerge at the quantum level with a finite one-loop coefficient 
and as two-loop (scale-invariant) counterterms, due to the non-renormalizability of 
theories with quantum scale invariance. The role of these  (scale invariant) effective
 operators which capture the effects of an infinite series of  polynomial operators 
  deserves further study.

The quantum consistency of the calculation was verified by showing
that the Callan-Symanzik equation of the quantum potential is respected  
 in the presence of the non-polynomial operators.
 We also  showed the differences between the scale-invariant one-loop potential 
and its counterpart computed in the ``usual'' DR scheme ($\mu=$constant) that breaks 
scale  symmetry explicitly,  in the presence at the tree level  of  
non-polynomial operators.

In quantum scale invariant models all mass scales are
generated by vacuum expectation
values of the fields, after spontaneous scale symmetry breaking; 
therefore, any mass hierarchy is not primary or fundamental, but can be generated by  
a hierarchy of the  (dimensionless) couplings of  the theory. 
The vacuum energy is vanishing at the loop level
in  the case of  spontaneously broken quantum scale symmetry
provided one coupling is a  function of the rest; this ensures 
the flat direction exists.
This can be arranged by one initial  classical tuning, with subsequent, quantum
 tunings  of $\cO(\lambda_j)$ relative to previous order.
This picture is in contrast to the case when the regularization
 breaks explicitly the classical scale symmetry of the action, 
 leading to a different quantum theory
(where the minimum of  the  potential is non-zero).

It is possible to arrange  a hierarchy $m_{\tilde\phi}^2\sim
\langle\phi\rangle^2\ll\langle\sigma\rangle^2$
by choosing at the classical level an ultraweak coupling $\lambda_m$ between the SM and the
 hidden sector of the dilaton ($\vert\lambda_m\vert\ll~\lambda_\phi$)
or by  more elegant
 means (dynamics, etc). This hierarchy is  stable at the one-loop level, 
without additional tuning of the couplings and despite the presence of the
non-renormalizable operators mentioned. This UV behaviour should survive to higher orders
due to  the spontaneous (i.e. soft) scale symmetry breaking.

\bigskip
\section*{Appendix}

\def\theequation{A-\arabic{equation}}
\def\thesubsection{A}
\setcounter{equation}{0}
 \label{appendixA}

For convenience, we present the expressions of the beta functions found in the text
\bea
\beta_{\lambda_\phi}&=&
\frac{1}{\kappa}
\,\Big[\,
3\, \Big(\frac{9}{4} g_2^4+ \frac{3}{4} g_1^4 +\frac{3}{2} g_1^2 g_2^2-12 h_t^4\Big)
\nonumber\\[-3pt]
&- &
4\lambda_\phi \,\Big(\frac{3}{4} g_1^2 +\frac{9}{4} g_2^2 - 3 h_t^2\Big)
+4\lambda_\phi^2 
+3\lambda_m^2+96 \lambda_m\lambda_6\Big],
\nonumber
\nonumber\\[-3pt]
\beta_{\lambda_m}&=&
\frac{2\lambda_m}{\kappa}
\Big[
\lambda_\phi+2\lambda_m +\frac{1}{2}\lambda_\sigma -\Big(\frac{3}{4} g_1^2+\frac{9}{4} g_2^2 -3 h_t^2\Big)\Big],
\nonumber\\[-3pt]
\beta_{\lambda_\sigma}&=&
\frac{3\lambda_\sigma}{\kappa} \Big[ \lambda_\sigma+4\frac{\lambda_m^2}{\lambda_\sigma}\Big],
\nonumber\\[-3pt]
\beta_{\lambda_6}&=&\frac{3\lambda_6}{\kappa}
\Big[ 6\lambda_\phi-8\lambda_m +\lambda_\sigma -2
 \Big(\frac{3}{4} g_1^2+\frac{9}{4} g_2^2 -3 h_t^2\Big)\Big],
\nonumber\\[-3pt]
\beta_{\lambda_8} &=&
\frac{2}{\kappa}
\Big[2\lambda_6 \,(28 \lambda_6+\lambda_m)-4\lambda_8
 \Big(\frac{3}{4} g_1^2+\frac{9}{4} g_2^2 -3 h_t^2\Big)\Big],
\nonumber\\[-3pt]
\beta_{\lambda_{10}}&=&
\frac{10}{\kappa}
\Big[ 4\lambda_6^2 -\lambda_{10} \Big(\frac{3}{4} g_1^2+\frac{9}{4} g_2^2 -3 h_t^2\Big)\Big],
\nonumber\\[-3pt]
\beta_{\lambda_{12}}&=&
\frac{2}{\kappa}
\Big[ 3\lambda_6^2 -6\lambda_{12} \Big(\frac{3}{4} g_1^2+\frac{9}{4} g_2^2 -3 h_t^2\Big)\Big].
\eea

\bigskip\bigskip
\bigskip\noindent
{\bf Acknowledgements:    }
The  work of D.M.G.  was supported by a grant of the Romanian National  Authority for
Scientific Research (CNCS-UEFISCDI) under  project number PN-II-ID-PCE-2011-3-0607.
The work of Z.L. was supported by the National Science Centre under research 
grant DEC-2012/04/A/ST2/00099.  The work of P.O. was supported by the
National Science Centre, Poland, under research grant DEC-2016/21/N/ST2/03312.

\end{document}